# On the importance of constrained atomic relaxations in the Nudged Elastic Band calculations of the Peierls barriers of dislocations

R. Gröger[1,*] and V. Vitek[2]

[1] Central European Institute of Technology - Institute of Physics of Materials, Academy of Sciences of the Czech Republic (CEITEC-IPM), Žižkova 22, 61662 Brno, Czech Republic
[2] Department of Materials Science and Engineering, University of Pennsylvania, Philadelphia, PA 19104, USA

**Abstract**

We demonstrate that the straightforward application of the Nudged Elastic Band (NEB) method does not determine the correct Peierls barrier of $1/2\langle 111\rangle$ screw dislocations in BCC metals. Although this method guarantees that the states (images) of the system are distributed uniformly along the minimum energy path, it does not imply that the dislocation positions are distributed uniformly along this path. In fact, clustering of dislocation positions near potential minima occurs which leads to an overestimate of both the slope of the Peierls barrier and the Peierls stress. We propose a modification in which the NEB method is applied only to a small number of degrees of freedom that determine the position of the dislocation, while all other coordinates of atoms are relaxed by molecular statics as in any atomistic study. This modified NEB method with relaxations gives the Peierls barrier that increases smoothly with the dislocation position and the corresponding Peierls stress agrees well with that evaluated by the direct application of stress in the atomistic modeling of the dislocation glide.

* Corresponding author.  E-mail: groger@ipm.cz

# 1  Introduction

One of the most important problems in physics, chemistry and materials science is to find the minimum energy path of a system of particles between its initial and final state during processes such as thermally activated transitions. If the state of a system composed of *N* particles is described by their coordinates, its evolution can be represented by a trajectory in the *3N* dimensional configurational space spanned by the coordinates of all particles. The main problem is then stated as follows: "Provided the initial and final states of the system are known and assuming that we can calculate the potential energy of the system at every point in the configurational space, determine the path through this space that minimizes the potential energy expended by the system in transition from the initial to the final state." In the continuum limit, the total energy used by the system in such transition is equal to the line integral of the potential energy along a path through the configurational space between two known fixed states. Each state along this chain of states is represented by a *3N*-dimensional vector $\mathbf{r} = [\mathbf{r}_1 \quad \mathbf{r}_2 \quad ... \quad \mathbf{r}_N]$, where $\mathbf{r}_i = [x_i \quad y_i \quad ... \quad z_i]$ is the position vector of the atom *i* in the three dimensional space. When relaxation of atomic positions is allowed the states evolve such that their displacements are proportional to the force $-\nabla E(\mathbf{r}) = [-\nabla E(\mathbf{r}_1) \quad -\nabla E(\mathbf{r}_2) \quad ... \quad -\nabla E(\mathbf{r}_N)]$ resulting from the interaction between particles, which we shall call hereafter the potential force. The numerical implementation of this procedure begins by discretizing the path by a finite number of intermediate states that are obtained by interpolating the positions of atoms between the two fixed states. Each of these states moves in the configurational space towards the position where $\nabla E(\mathbf{r}) = \mathbf{0}$. This represents a major problem since in this case all the states will be found only at potential extrema, where $\nabla E(\mathbf{r}) = \mathbf{0}$, and the path of the system through the regions $\nabla E(\mathbf{r}) \neq \mathbf{0}$ will not be resolved.

A remedy of this problem was offered with the introduction of the now popular Nudged Elastic Band (NEB) method (Henkelman, Jóhannesson and Jónsson 2000, Henkelman and Jónsson 2000, Henkelman, Uberuaga and Jónsson 2000, Jónsson, Mills and Jacobsen 1998). Within this method, one connects each pair of neighboring states of the system (images) in the *3N* dimensional configurational space by a linear spring and replaces the parallel component of the potential force by the parallel component of this spring force. The force on each image is thus defined as a vector sum of the component of the potential force perpendicular to the path and the component of the spring force parallel to the path. If all springs have the same stiffness, the states of the system are distributed uniformly along the path of the system through the *3N* dimensional configurational space, which avoids clustering of the states at the minima of $E(\mathbf{r})$. The cross-section of the potential energy hypersurface along this path provides both the shape and the height of the barrier that has to be overcome by the system in order to transfer from the initial to the final state. This procedure requires calculations of the potential energy of an ensemble of atoms that interact via the corresponding interatomic forces. These are typically obtained using interatomic potentials or the Density Functional Theory (DFT).

In this paper, we show that the NEB method has to be significantly modified when calculating the Peierls barriers of dislocations. We introduce this modification for the $1/2\langle 111 \rangle$ screw dislocation in BCC metals, which is known to govern the low-temperature plastic deformation of these metals owing to the very high Peierls stress associated with its non-planar core. We first review the current formulation of the NEB method and show that the equal spacing of images in the $3N$ dimensional configurational space does not imply a uniform spacing of the dislocation positions along the path. Instead, a clustering of the dislocation positions around the minima of the barrier occurs, while still maintaining the uniform distribution of states in the $3N$ dimensional configurational space. This leads to a significant overestimate of the slope of the Peierls barrier and, therefore, also the magnitude of the Peierls stress deduced from it. In order to mitigate this problem, we propose a modification whereby the NEB method is applied only to a small number, $D$, of degrees of freedom (marked DOFs hereafter) of the system that describe closely the position of the dislocation, while the remaining *3N-D* DOFs are fully relaxed using a molecular statics energy minimization. This constitutes a severe reduction of the dimensionality of the space in which the NEB method operates, typically 2-3 orders of magnitude in the case of empirical potentials. In order to confirm the necessity to modify the NEB method, we demonstrate that this procedure yields the Peierls stress that agrees well with the results of the atomistic studies of the dislocation motion under the applied stress (Gröger, Bailey and Vitek 2008) whilst the standard NEB method overestimates the Peierls stress significantly.

## 2   Interatomic potential and the geometry of the simulated block

All calculations in this paper were carried out using the Bond Order Potential (BOP) that was developed for body-centered cubic (BCC) tungsten in by Mrovec, Gröger, Bailey, Nguyen-Manh, Elässser and Vitek (2007). The simulated block of atoms is oriented such that the *x* axis is parallel to the $[\bar{1}2\bar{1}]$ direction, the *y* axis to the $[\bar{1}01]$ direction, and the *z* axis to the [111] direction and thus also to the line of the dislocation. The block is disc-shaped, with three (111) planes stacked along the $z$ axis of the disc with interplanar separations equal to $b/3$, where $b$ is the magnitude of the Burgers vector of the dislocation. Periodic boundary conditions are applied along the *z* axis to model an infinite straight dislocation. The radius of the block in the (*x,y*) plane is $R = 21a$, where $a = 3.1652$ Å is the lattice parameter. The initial and final configurations that correspond to the dislocation positioned in two neighboring sites in the $(\bar{1}01)$ slip plane are obtained by first displacing all 2406 atoms in the block according to the anisotropic elastic field of the dislocation (Hirth and Lothe 1982), and then relaxing 1221 atoms inside the region of $R/a \leq 15$ while keeping the remaining atoms at the positions determined by anisotropic elastic displacements. The relaxation was terminated when all forces on atoms in the block were less[1] than $0.001 \text{ eV/Å}$. These two configurations are shown in Fig. 1, where circles represent atoms in three successive (111) planes perpendicular to the dislocation line. The arrows are proportional to the [111] component of the relative displacements of the neighboring atoms produced by the dislocation (Vitek, Perrin and Bowen 1974). In Fig. 1a (initial state), the center of the dislocation is identified with the midpoint of the triangle of atoms 1, 4 and 3, while in Fig. 1b (final state) it is identified with the midpoint of the triangle of atoms marked 2, 5 and 4. The actual path of the dislocation between these two positions is an unknown curve and the goal of NEB-type calculations is to determine this path.

---

[1] This ensures that the energies of all movable images obtained by the NEB calculation are higher than those of the two fixed images.

[ Insert Fig. 1 about here ]

## 3  Calculation of the Peierls barrier using the standard NEB method

When a dislocation moves in the crystal, it experiences a periodic lattice friction determined by the Peierls barrier. In order to investigate the temperature dependence of the dislocation glide that occurs via formation and propagation of pairs of kinks (Caillard and Martin 2003, Dorn and Rajnak 1964), one needs to know not only the height of this barrier but also its overall shape. There have been several attempts to calculate the Peierls barriers for $1/2\langle 111 \rangle$ screw dislocations in BCC metals – both using the NEB method (Rodney and Proville 2009, Ventelon and Willaime 2007) and by first constructing a two-dimensional Peierls potential based on atomistic studies of the glide of the screw dislocation under stress at 0 K and then searching for the minimum energy path of the dislocation in this potential (Gröger and Vitek 2008). In this section we investigate the determination of the Peierls barrier using the standard NEB approach.

Suppose the path of the dislocation is discretized by a finite number $M$ of states (images), where each of the states is fully determined by the $3N$ coordinates of the atoms of the block. Based on knowledge of the initial and final states of the system, $I = 0$ and $M + 1$, the NEB method finds a chain of states $I = 1...M$ for which the total energy of the system is minimized. The evolution of the chain of states towards this minimum energy path is driven by the NEB force

$$\mathbf{F}_{NEB}^I = -\nabla_\perp E^I + \mathbf{F}_{s\|}^I \qquad (1)$$

that determines displacements of all atoms in the image $I$. Here, $-\nabla_\perp E^I$ is the component of the potential force ($-\nabla E^I$) in the direction perpendicular to the path, defined in the Introduction, and $\mathbf{F}_{s\|}^I$ the component of the spring force parallel to the path. The former arises due to the variation of the energy of the system with the state $\mathbf{r}^I$ of the image $I$ in the $3N$ dimensional configurational space. After displacing the atoms in all movable states $I = 1...M$, the energies of these new configurations and forces on atoms are recalculated. The neighboring states are connected by linear springs with a spring constant $k$. We used $k = 50$ eV/Å$^2$ that represents the largest spring constant for which the steepest descent minimization using the NEB force (1) remains stable. We accept the chain of states as a representation of the minimum energy path when $\max_{I=1...M} |\nabla_\perp E^I| < 0.01$ eV/Å and, simultaneously, $\max_{I=1...M+1} \left[ \Delta^2_{3N(I-1,I)} - \overline{\Delta}^2_{3N} \right] < 0.0005$ Å$^2$. Here, $\Delta^2_{3N(I-1,I)}$ is the square of the distance between neighboring images $I - 1$ and $I$, and $\overline{\Delta}^2_{3N}$ the mean squared spacing of neighboring images, both measured in the $3N$ dimensional configurational space. The latter criterion is more convenient than a criterion using the spring force directly because in this case the termination criterion does not depend on the magnitude of the spring constant. The energy per unit length of a straight dislocation at different points along the path is then calculated as

$$V^I = \frac{E^I - E^0}{b} ,\qquad(2)$$

where $E^I$ is the total energy of the atomic block that represents the image $I$, and $b$ the Burgers vector.

In the following, we define the transition coordinate $\xi$ as the distance of the dislocation from its position in the image $I = 0$, measured along the path of the dislocation. Unfortunately, the position of the dislocation cannot be determined unambiguously from the coordinates of all atoms in the block and thus also the path of the intersection of the dislocation line with the (111) plane is not known unequivocally. In this paper, we approximate the position of the dislocation in the image $I$ as $\xi^I = [I/(N+1)]a_0$, where $a_0 = a\sqrt{2/3}$ is the shortest distance between the two neighboring minimum energy lattice sites, and $a$ the lattice parameter. This assumption of a straight path of the dislocation was adopted also by Rodney and Proville (2009), and Ventelon and Willaime (2007). The variation of the energy of the dislocation, Eq. 2, along this transition coordinate is plotted in Fig. 2 in black. The same barrier has been obtained when the path of the dislocation was discretized using $M$ = 5, 15 and 32 images. The barrier increases slowly at first but then rises to its maximum rapidly. Clearly, this abrupt change of the slope calls for a deeper understanding.

[ Insert Fig. 2 about here ]

There are several underlying assumptions within the NEB method that may affect the trajectory along which the chain of states moves towards the minimum energy path. First, no additional atomic movements take place within the NEB method over and above those controlled by the force given by Eq. (1) and, therefore, all DOFs are treated on equal footing. In order to sample the pathway at the points of the potential energy hypersurface where $\nabla E \neq \mathbf{0}$, the NEB method replaces the parallel component of the potential force by the parallel component of the spring force. Second, all springs connecting images have typically the same spring constants, although some simple rules were proposed that relate the stiffnesses of springs to the energies associated with images (Henkelman, Uberuaga and Jónsson 2000). This is in contrast to the original problem, where the differences in parallel components of the potential force $-\nabla_\parallel E$ of two neighboring images vary with the position of these images in the configurational space. To the best of our knowledge, the validity of these assumptions has never been investigated and questioned, which is our objective in the following.

As mentioned above, the presence of linear springs in the NEB method guarantees that the states (images) are spaced uniformly along the path through the *3N* dimensional configurational space. The square of the distance between two neighboring states $I-1$ and $I$ in this space is

$$\Delta^2_{3N(I-1,I)} = \sum_{i=1}^{3N} \left(r_i^I - r_i^{I-1}\right)^2 ,\qquad(3)$$

where $r_1^I...r_{3N}^I$ represent the 3*N* DOFs of the state *I* that have been found when employing the NEB method. No matter what the magnitude of the spring constant, the assumption of a uniform distribution of states along the path means that at the end of the calculation $\Delta^2_{3N(I-1,I)} = \Delta^2_{3N(J-1,J)}$ for any pair of states $I, J \in \langle 1, M+1 \rangle$. This is shown in Fig. 3, where we plot as the solid black line the variation of the separations between neighboring images in the 3*N* dimensional configurational space. Obviously, the value of $\Delta^2_{3N}$ is constant along the chain. The position of the dislocation, of course, changes as one traces the path of the system in the configurational space. Owing to the uniform spacing of the images in the *3N* dimensional space, it is tempting to conclude that the positions of the dislocation embedded in individual images are also distributed uniformly along this path.

[ Insert Fig. 3 about here ]

In order to test the validity of this conjecture, we need to determine for every image *I* the position of the dislocation. This is a very difficult task when considering all the atoms in the block. However, the position of the dislocation in the lattice is fully determined if one knows the displacements of atoms close to the center of the dislocation. In other words, if the positions of these atoms are known and are held fixed, then the displacements of the remaining atoms farther away from the dislocation center do not alter the position of the dislocation. This implies that the coordinates of atoms in every image *I* can be divided into two groups. A small number of DOFs, *D*, can be chosen that determine the position of the dislocation. The remaining DOFs, the number of which is $3N - D$ that is typically much larger than *D*, do not affect the position of the dislocation. Hence, we can write Eq. (3) as $\Delta^2_{3N(I-1,I)} = \Delta^2_{D(I-1,I)} + \Delta^2_{3N-D(I-1,I)}$, where $\Delta^2_{D(I-1,I)}$ is the square of the distance between states $I-1$ and *I* measured in the subspace of the dimension *D* and, similarly, $\Delta^2_{3N-D(I-1,I)}$ is the squared distance measured in the complementary subspace of the dimension $3N - D$. The conjecture above can now be restated as: "The uniform spacing of images in the 3*N* dimensional configurational space implies their uniform spacing also in the subspace of dimension *D*." In order to see whether this statement is true, we recall Figs. 1a,b that show the initial and final states of the system. As an example, we consider the atoms marked by the numbers 1-5 to be the closest to the path of the dislocation between the two states in the $(\bar{1}01)$ slip plane and the $[\bar{1}2\bar{1}]$ direction. Moreover, since the dislocation is of the screw character, only the *z* coordinates of these atoms need to be known to determine the position of the dislocation at any point along the path. Hence, we define the subspace of the dimension $D = 5$ that is spanned by the *z* coordinates of these five atoms. The square of the spacing of images in this subspace, $\Delta^2_D$, is obtained similarly as in Eq. (3), with the exception that 3*N* is now replaced by $D = 5$ and the sum is only over the DOFs that span this *D* dimensional subspace. The variation of the spacing $\Delta^2_5$ for the minimum energy path that was obtained by the standard NEB method is shown in Fig. 3 by the dashed black curve. It is clearly not uniform, which disproves the conjecture above. Hence, we have shown that the uniform spacing of images in the 3*N* dimensional configurational space does not imply a uniform spacing of the dislocation positions along this path. Consequently, we can conclude that the standard NEB method with *3N* DOFs does not converge to the correct minimum energy path of the dislocation and, therefore, the barrier defined by this path is not the correct Peierls barrier of the dislocation.

## 4 NEB method with constrained atomic relaxations

In order to find the minimum energy path along which the spacing of the dislocation positions is uniform, we propose to apply the NEB method only to the $D$ DOFs, where $D \ll 3N$. As an example we take $D = 5$, where the DOFs correspond to the *z* coordinates of the atoms 1-5 in Fig. 1a,b. All other DOFs, i.e. the *x* and *y* coordinates of these five atoms and (*x*,*y*,*z*) coordinates of all other atoms in the block that are not included in the NEB calculation are adjusted using the lattice statics relaxation that minimizes the energy of the block during which the *z* coordinates of the five chosen atoms are held fixed. The energy associated with the image $I$ then represents a potential energy of a *relaxed* block of atoms with the dislocation at a well-defined position. This is in contrast to the application of the NEB method presented in Section 3, where no atomic relaxation is done and the energy of every image is obtained using the coordinates of all $N$ atoms as they evolve according to Eq. (1) and thus follow the standard NEB path. To distinguish our modification from the standard NEB method, we propose to call it the NEB method with relaxations and use a shorthand notation NEB+r.

The variation of the energy of the dislocation along the path, obtained by the NEB+r method for $D = 5$, is shown in Fig. 2 by red symbols. One can clearly see that the top of this barrier is the same as that found using the standard NEB method. This means that both the NEB and the NEB+r methods predict the paths of the system that pass through the saddle-point of the potential energy hypersurface, $E(\mathbf{r})$. However, unlike in the standard NEB method, the barrier obtained by the NEB+r method increases smoothly towards its maximum, without any cusps or sudden changes of the slope of the barrier. The spacings between neighboring images obtained using the NEB+r method are shown in Fig. 3 in red. As expected, the red dashed line marked $\Delta_5^2$ (NEB+r) in Fig. 3 shows that the images, and thus also the positions of the dislocation, are distributed uniformly along the path. Moreover, the method automatically yields a uniform separation of images in the complementary subspace of dimension $3N - 5$ (red dotted line) as well as in the entire configurational space of dimension $3N$ (red solid line), although no constraint was imposed on the system to do so. We can thus state the following: "Within the NEB+r method, the uniform spacing of images in the $D$ dimensional subspace that is spanned by the *z* coordinates of the atoms closest to the path of the dislocation center implies equal spacing of the images in the full $3N$ dimensional configurational space." However, the squared spacing of the images $\Delta_{3N}^2$, obtained from the standard NEB calculation, is about 0.021 Å$^2$, while that obtained from the NEB+r calculation only 0.014 Å$^2$ (see the solid lines in Fig. 3). Taking into account the fact that the chain of states is made up of 16 linear segments implies that the path of the system obtained from the NEB+r method is about 0.43 Å ($\approx 0.17 a_0$) shorter than that obtained from the standard NEB method, while both paths pass through the same saddle-point of $E(\mathbf{r})$.

[ Insert Fig. 4 about here ]

In the calculations discussed in this section, only the $z$ coordinates of the atoms marked 1-5 in Fig. 1 define the position of the dislocation. Hence, it is important to investigate how these five DOFs (and thus the position of the dislocation) change as the path of the system evolves towards the minimum energy path. In every iteration of the standard NEB method, the position $\xi^I$ of an image $I$ along the instantaneous path of the system through the subspace of the dimension $D = 5$ is

$$\xi^I = \xi^{I-1} + \Delta_{D(I-1,I)} , \qquad (4)$$

where $\xi^0 = 0$ (initial state). The variation of $\Delta^2_{D(I-1,I)}$ along the chain of states is plotted in Fig. 3 by the dashed black line and in Fig. 4 the curves show how the positions $\xi^I$ of individual images evolve during the NEB calculation. Initially (iteration 0), the images are distributed uniformly along the path. As the calculation proceeds, the red and blue curves cluster towards those corresponding to the two fixed states. Taking into account that the five chosen DOFs represent the position of the dislocation, we conclude that the dislocation positions cluster towards the initial and final states of the system, respectively. We checked that this becomes more prominent as the number of images along the chain increases and/or when the size of the system, i. e. the number of atoms $N$, increases. Moreover, Fig. 4 also shows that the position of the last image ($I = 16$) moves away from the position of the dislocation in the first image ($I = 0$) and, therefore, the length of the path increases.

[ Insert Fig. 5 about here ]

In all calculations using the NEB+r method, we adjusted by the NEB force only the $z$ coordinates of the five atoms that are closest to the path of the dislocation (see Fig. 1). In order to assure that our conclusions are not dependent on this choice, we have investigated possible changes of the Peierls barrier when more than 5 DOFs are included. In Fig. 5a, we show three regions of atoms that contain 5, 11 and 17 atoms, respectively. The innermost region contains the five atoms that define the center of the dislocation as it moves between the two neighboring minimum energy lattice sites. If all $x$, $y$, and $z$ coordinates of these atoms are taken as DOFs of the NEB+r calculation, there are 15, 48 and 99 DOFs enclosed by the boundaries of the three regions. In Fig. 5b, we show the Peierls barriers obtained using the NEB+r method in which 48 and 99 DOFs were employed, together with the results of the NEB+r calculation using 5 DOFs and the standard NEB calculation with $3N$ DOFs. Only one half of the barrier is shown owing to the symmetry. The barrier obtained using 48 DOFs (dashed green curve) is practically the same as that obtained using 5 DOFs (red solid line). A visible, albeit still small, deviation towards the barrier obtained using the standard NEB method occurs in NEB+r calculation with 99 DOFs (blue dashed line in Fig. 5b). Hence, the Peierls barrier obtained by the NEB+r method depends only weakly on the number of DOFs employed in the evaluation of the NEB force and this provides further proof of the credibility of the NEB+r method.

The NEB+r calculation may seem to be more computationally demanding than the straightforward application of the NEB method because it requires the minimization of the energies of all $M$ images at each step. However, this is not necessarily the case, because the number of the NEB+r steps needed to achieve an acceptable approximation of the transition path decreases by about an order of magnitude. This is further underscored by the fact that even though the gradient of the potential energy $-\nabla E$ points in the direction of the steepest descent of the energy, this may not be the case for its projection into the direction perpendicular to the path, $-\nabla_\perp E$, that is used to define the NEB force (1). For complex potential energy hypersurfaces, the latter can even cause the image to move uphill, which disturbs the convergence of the chain of states towards the minimum energy path. We observed that these uphill movements are more frequent as the size of the simulated block increases, and thus also the complexity of the configurational space increases. None of these artifacts are present when the NEB+r method is used to calculate the minimum energy path. The reason is that it operates in the subspace of dimension $D$, which is significantly less complex than the full $3N$ dimensional configurational space. The chain of states then converges towards the minimum energy path without any transfer of the images to higher energies.

## 5 Comparison of the Peierls stress obtained from the NEB and NEB+r methods with the direct application of stress in the atomistic model

It is important to compare the Peierls stresses that are predicted by the barriers in Fig. 2 with the value obtained by the direct application of stress in the atomistic model of the glide of the dislocation. The Peierls stress $\sigma_P$ is related to the maximum slope of the Peierls barrier as $\sigma_P b = \max(dV/d\xi)$. From Fig. 2, this yields $\sigma_P/C_{44} = 0.034$ for the standard NEB method and 0.025 for the NEB+r method, while $\sigma_P/C_{44} = 0.027$ was found in the atomistic modeling of the screw dislocation glide under the applied shear stress in the slip direction[2]. The value obtained from the NEB+r methods deviates by 8% from the value obtained atomistically, while that obtained from the standard NEB method deviates by 26% from this value. Consequently, the Peierls stress is very close to the value obtained by differentiating the Peierls barrier obtained from the NEB+r method while it is significantly overestimated by the standard NEB method.

Nevertheless, it could be argued that the deviation of the latter from the value obtained in the atomistic study would be reduced if the actual curved path of the dislocation were correctly taken into account since the path between the two fixed states would be longer and thus the gradient $dV/d\xi$ would be lower. This would lower the Peierls barriers obtained from both NEB methods and the one obtained from the standard NEB method could be in a better agreement with the direct application of stress than that obtained using the NEB+r method. However, this argument is incorrect. If the path of the dislocation deviates from the straight line connecting two neighboring minimum energy sites, the relation between the shape of the Peierls barrier and the magnitude of the Peierls stress has to be generalized so that $\sigma_P b \cos\psi = \max(dV/d\xi)$, where $\psi$ is the angle between the tangent to the path of the dislocation and the $(\bar{1}01)$ plane. As a result, if a curved path of the dislocation is considered, *both* the maximum slope of the Peierls barrier *and* the projection of the Peierls stress into the direction of the path decrease simultaneously. For this reason our conclusion regarding the agreement between the Peierls stresses obtained by differentiating the barriers and the direct application of the stress in the atomistic model holds even in the case of a curved dislocation path.

---

[2] In Section 2, we relaxed the two fixed images down to the maximum force on atom $F_{max} < 0.001\,\mathrm{eV/\mathring{A}}$. To make a consistent comparison of the Peierls stress obtained from the Peierls barrier with that calculated by the direct application of stress, the atomic configurations in the latter have to be determined with the same precision, i. e. the same $F_{max}$. In our previous studies (Gröger, Bailey and Vitek 2008), we considered $F_{max} < 0.005\,\mathrm{eV/\mathring{A}}$, for which we obtained the Peierls stress $\sigma_P/C_{44} = 0.028$. We now performed a new calculation of the Peierls stress by a direct application of stress, where $F_{max} < 0.001\,\mathrm{eV/\mathring{A}}$. From this calculation, $\sigma_P/C_{44} = 0.027$. We expect that even a slightly lower value of the $\sigma_P$ would be obtained if the relaxation could be continued towards $F_{max} = 0$.

## 6 Conclusions

We have demonstrated that the straightforward application of the NEB method does not lead to the correct Peierls barrier of the $1/2\langle 111\rangle$ screw dislocation in BCC metals. The reason is that a uniform spacing of images along the minimum energy path of the system in the $3N$ dimensional configurational space, where $N$ is the number of atoms in the studied block, does not imply a uniform spacing of the dislocation positions throughout these states. On the contrary, we have demonstrated that the spacing of the dislocation positions along the path of the system is strongly non-uniform, with more positions close to the minima of the barrier and fewer close to its maximum.

In order to circumvent this drawback, we introduced a modification in which the NEB method is applied only to a small number of DOFs that determine the position of the dislocation, while all the other DOFs are relaxed via standard molecular statics using the appropriate interatomic potential. This represents a severe reduction of the dimensionality of the problem from $3N$ (typically thousands) to as few as 5, while maintaining a uniform spacing of the dislocation positions along the minimum energy path. The Peierls barrier obtained using this modified NEB method with relaxations (NEB+r) gives the Peierls stress that agrees to within 8% with that obtained in the atomistic study of the dislocation glide under the applied shear stress. The NEB+r method can be used efficiently to investigate the variation of the Peierls barrier with stress that was found to be essential in atomistic studies of application of more complex loading than pure shear (Gröger, Bailey and Vitek 2008, Gröger and Vitek 2009). The superiority of the NEB+r method over the standard NEB method when calculating the Peierls barrier will obviously apply not only to $1/2\langle 111\rangle$ screws in BCC metals but to dislocations of any orientation in various structures. Moreover, the NEB+r method might be superior even more generally whenever long-range strain fields are associated with the defects involved in some transitions or transformations and/or when the activation energy associated with a thermally activated process depends explicitly on the path of the system between the two fixed states. In addition to the dislocation glide, other examples are absorption or emission of dislocations by and from interfaces, or lattice trapping of cracks.

## Acknowledgments

RG acknowledges support from the Czech Science Foundation, Grant no. P204/10/0255 and from the Academy of Sciences of the Czech Republic, Research Project no. AV0Z20410507. VV was supported by the US Department of Energy, Office of Basic Energy Science, Grant no. DE-PG02-98ER45702.

# Figure captions

**Figure 1:** The initial (a) and final (b) configurations of the 1/2[111] screw dislocation used for the calculation of the Peierls barrier using the NEB method. The atoms are depicted as circles with the colors that distinguish the three successive (111) atomic planes. In both figures, the three atoms that encompass the center of the dislocation are numbered.

**Figure 2:** Variation of the energy of a straight dislocation along the minimum energy path between the fixed states at $\xi/a_0 = 0$ and 1, where $a_0$ is the distance of the two neighboring minimum energy lattice sites in the $(\bar{1}01)$ plane in the $[\bar{1}2\bar{1}]$ direction. The black data (curve) has been obtained by the standard NEB method that operates in the configurational space of the dimension $3N$. The red data (curve) have been obtained using the NEB+r method in which only the $z$ coordinates of the five atoms marked 1-5 in Fig. 1 were adjusted by the NEB force (1), while the remaining DOFs were obtained by atomic relaxations using the standard molecular statics. In both cases, the barrier is shown to be independent of the number of images used to discretize the path.

**Figure 3:** Distances between neighboring states obtained using the standard NEB method and the NEB+r method, calculated in the full configurational space of dimension $3N$ (solid lines), and two subspaces – of dimension $D = 5$ spanned by the $z$ coordinates of the atoms 1-5 in Fig. 1 (dashed lines), and in the complementary subspace of the dimension $3N - D$ (dotted lines).

**Figure 4:** Evolution of positions of images ($I$) obtained from the standard NEB method in the subspace of the dimension $D = 5$ spanned by the $z$ coordinates of the atoms 1-5 in Fig. 1. Each curve shows how the position of a particular image $I$ in this subspace varies during the calculation. Notice how the images $I = 1, 2$ (blue lines) and $I = 14, 15$ (red lines) cluster towards the two fixed states $I = 0$ and $I = 16$, respectively, as the calculation proceeds.

**Figure 5:** Dependence of the Peierls barrier obtained using the NEB+r method on the number of DOFs that are adjusted using the NEB force. (a) The atoms in the three shaded regions containing 5, 11 and 17 atoms enclose 15, 48 and 99 DOFs, respectively; these DOFs are adjusted according to the NEB force. The Peierls barriers for the latter two are shown in (b) together with the results of the NEB+r calculation using the 5 DOFs and the standard NEB calculation. In all calculations, the path was discretized by 15 images but only a half of the barrier is shown in (b) owing to the symmetry.

**Figures**

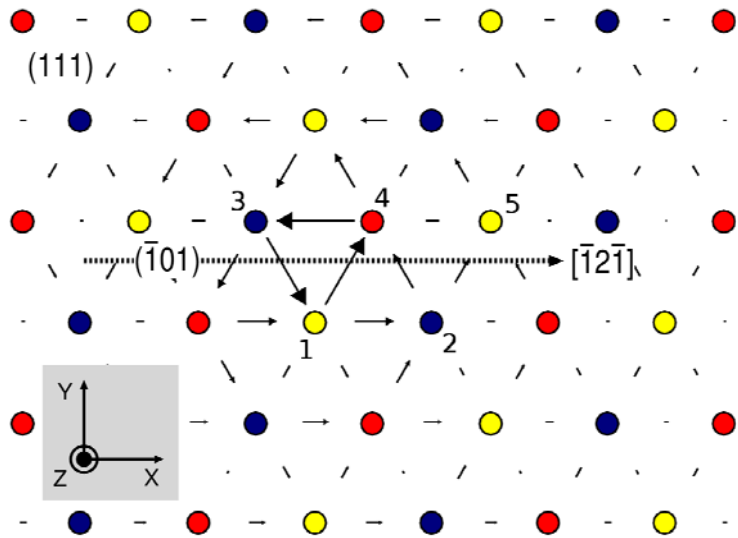

(a)

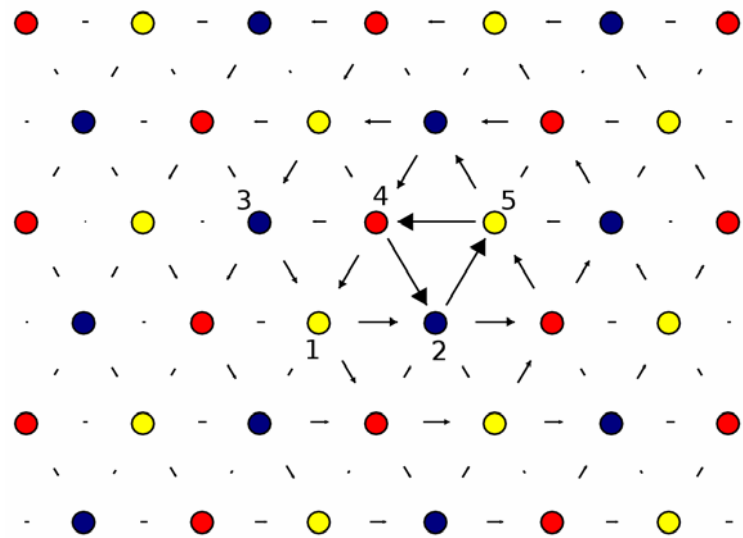

(b)

**Figure 1**

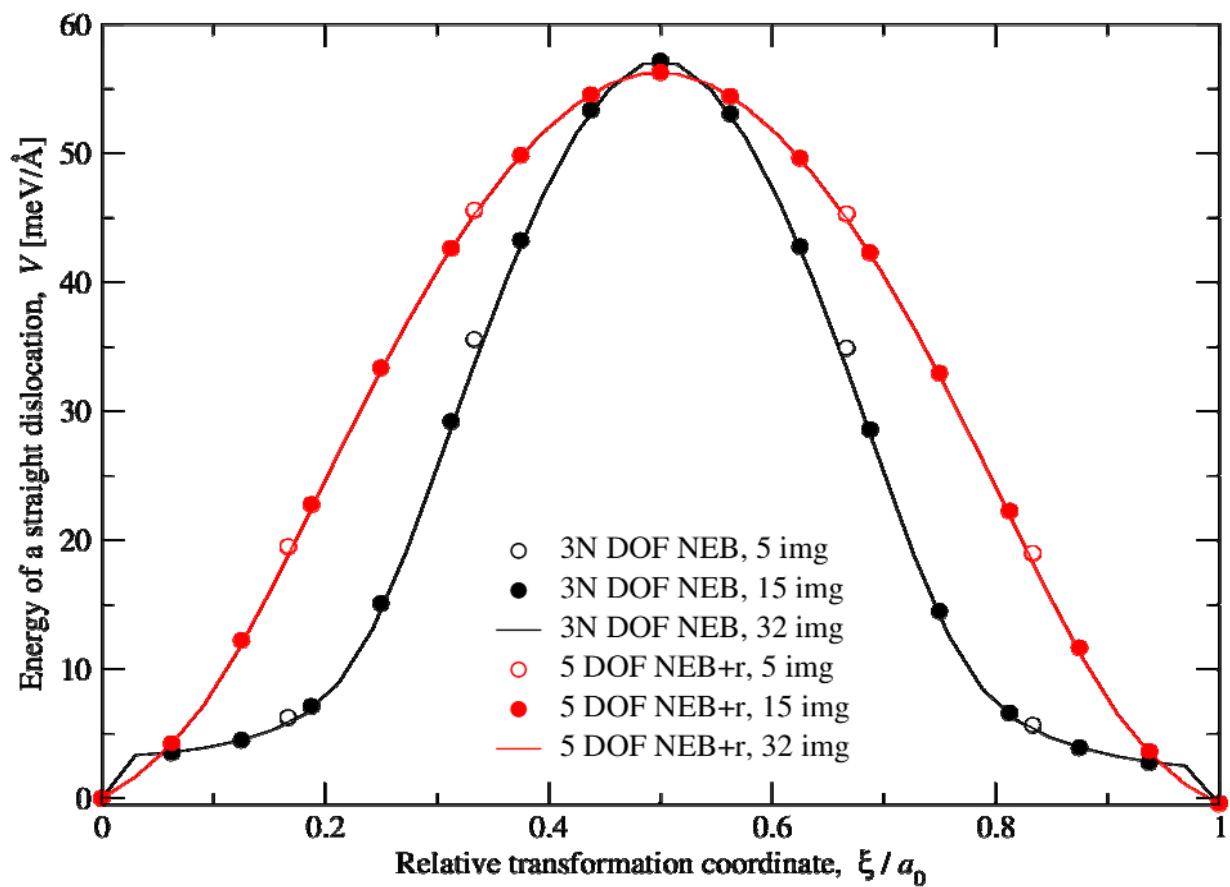

**Figure 2**

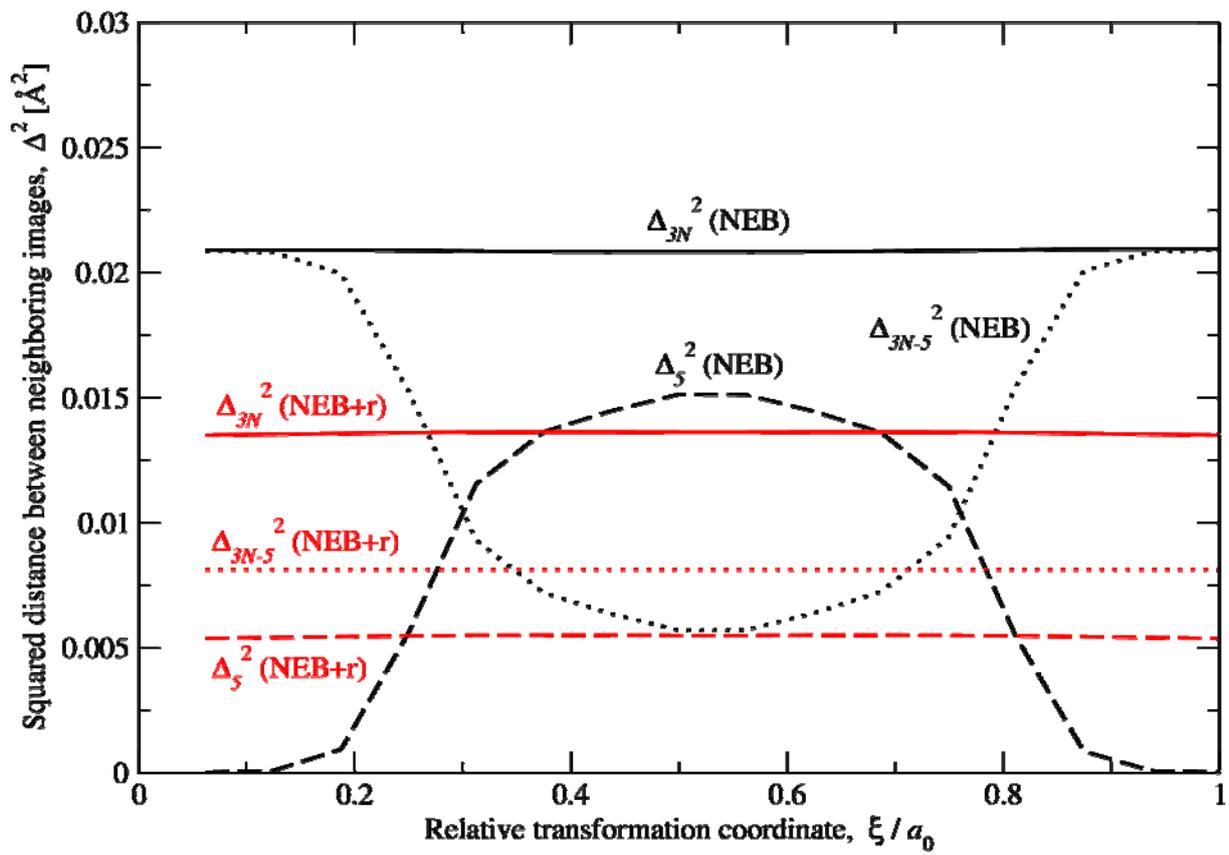

**Figure 3**

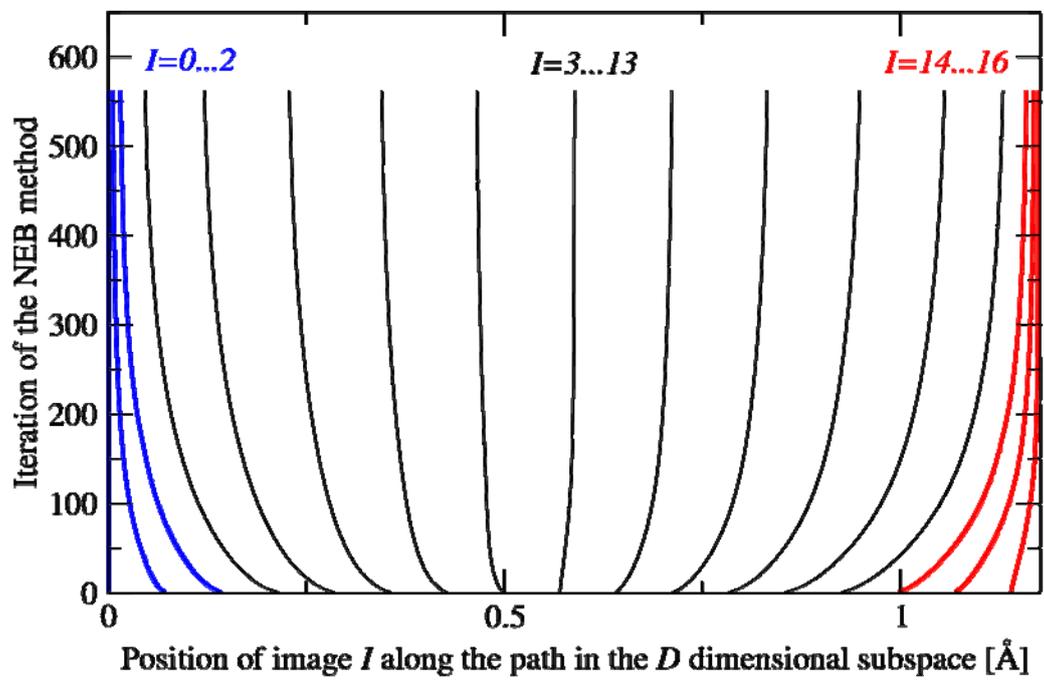

**Figure 4**

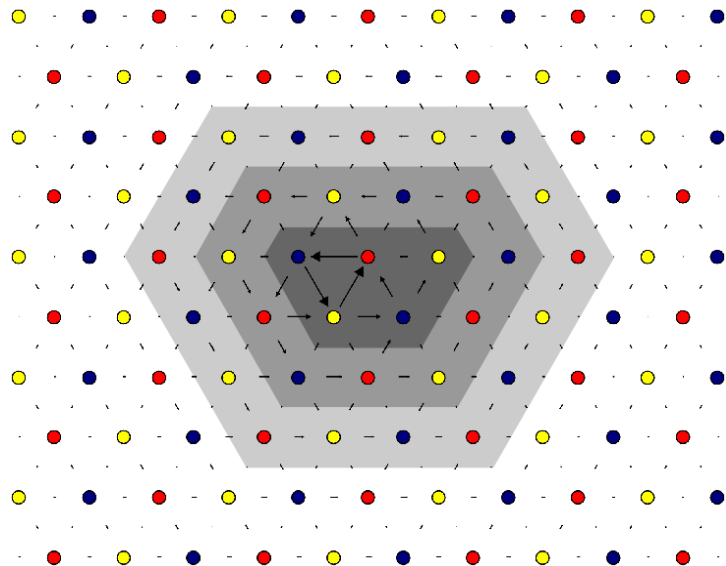

(a)

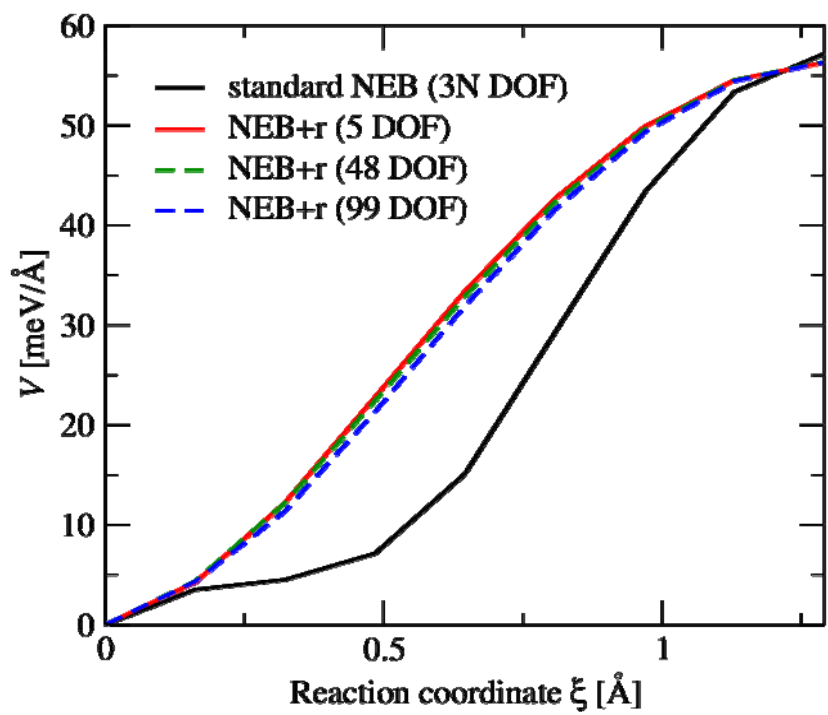

(b)

**Figure 5**